\documentclass[twocolumn,doublespacing,showpacs,superscriptaddress,amssymb,10pt]{revtex4}

\usepackage{graphicx}
\usepackage{dcolumn}
\usepackage{bm}
\usepackage{epsfig}
\usepackage{color}
\usepackage{longtable}
\usepackage{amsmath}
\usepackage{multirow}
\usepackage{tabularx}
\usepackage{pdfpages}

\definecolor{aogreen}{rgb}{0.0, 0.5, 0.0}

\def\ketm#1{  \left\vert  #1   \right\rangle   }

\def\sprm#1#2{  \left\langle #1 \left\vert \right. #2 \right\rangle   }

\def\mem#1#2#3{  \left\langle #1 \left\vert  #2 \right\vert #3 \right\rangle   }

\def\sixjm#1#2#3#4#5#6{  \left\{ \begin{array}{ccc}
                                               #1 & #2 & #3  \\
                                               #4 & #5 & #6
                     \end{array} \right\}   }
\definecolor{mymainmessagecolor}{RGB}{10,200,10}
\begin{document}

\preprint{}
\title{Photoelectron distribution of non-resonant two-photon ionization of neutral atoms}

\author{J.~Hofbrucker}
\affiliation{Helmholtz-Institut Jena, Fr\"o{}belstieg 3, D-07743 Jena, Germany}
\affiliation{Theoretisch-Physikalisches Institut, Friedrich-Schiller-Universit\"at Jena, Max-Wien-Platz 1, D-07743
Jena, Germany}%

\author{A.~V.~Volotka}
\affiliation{Helmholtz-Institut Jena, Fr\"o{}belstieg 3, D-07743 Jena, Germany}%

\author{S.~Fritzsche}
\affiliation{Helmholtz-Institut Jena, Fr\"o{}belstieg 3, D-07743 Jena, Germany}%
\affiliation{Theoretisch-Physikalisches Institut, Friedrich-Schiller-Universit\"at Jena, Max-Wien-Platz 1, D-07743
Jena, Germany}

\date{\today \\[0.3cm]}

\begin{abstract}

Photoelectron angular distributions following the non-resonant two-photon $K$-shell ionization of neutral atoms are studied theoretically. Using the independent particle approximation and relativistic second-order perturbation theory, the contributions of screening and relativistic effects to the photoelectron angular distribution are evaluated. A simple nonrelativistic expression is presented for the angle-differential cross section in dipole approximation for two-photon ionization by elliptically polarized photons, and its limitations are analyzed numerically. Moreover, we show that screening effects of the inactive electrons can significantly affect the photoelectron distributions and can also lead to a strong elliptical dichroism. Numerical results are presented for the case of two-photon $K$-shell ionization of neutral Ne, Ge, Xe, and U atoms. 

\end{abstract}

\newpage
\maketitle

\section{Introduction}
\label{Sec.Introduction}

Two-photon ionization is a fundamental nonlinear process which became an important benchmark for studying the interaction of intense light with matter. The study of nonlinear high-order processes has been driven by the recent development of free-electron lasers \cite{Pellegrini/RMP:2016}, intense-light sources operating in the ultraviolet and x-ray energy domains. In past, two-photon double ionization of Ne has been studied, and successfully detected at FLASH by performing a complete experiment \cite{Fritzsche/JPB:2008, Kurka/JPB:2009, Fritzsche/JPB:2009}. With today's high photon energies and beam intensities of free electron lasers, even two-photon ionization of the deep $K$-shell electron of medium and heavy elements became possible. In recent years, free-electron lasers have already been used to detect the two-photon $K$-shell ionization of neutral Ge, Cu, and Zr atoms \cite{Tamasaku/NP:2014, Szlachetko/SR:2016, Ghimire/PRA:2016}. In these experiments, the $K\alpha$ fluorescence was detected as a direct signature for producing a $K$-shell vacancy. However, two-photon ionization process can be also studied by a direct detection of the ejected photoelectrons. The total photoelectron yield has been measured in the case of two-photon ionization of the $4d$ subshell of Xe atom \cite{Richardson/PRL:2010}, while the angular distributions have been measured for the two-photon ionization of ground state helium \cite{Ma/JPB:2013, Mondal/PRA:2014}.

While the analysis of the total cross section enables one to obtain the amplitude ratios between the (two) dominant ionization channels, the photoelectron angular distribution additionally provides the information about their relative phases. 
Within nonrelativistic theory and dipole approximation, just these two atomic parameters are sufficient to fully characterize the photoelectron angular distribution \cite{Manakov/JETP:2009}. This approximate formulation, however, is expected to become insufficient for medium and heavy atoms, or when using high photon energies. The limitations of these approximations have been investigated especially in Refs. \cite{Koval/JPB:2004, Koval/Dissertation, Florescu/PRA:2012}, where two-photon ionization of H-like ions was considered. A different approach to calculate the amplitudes of two-photon ionization of H-like ions was taken in Ref. \cite{Staroselskaya/OSLP:2015}, where the variational method was used to calculate total as well as differential cross sections. Recently, the total cross section of two-photon ionization of xenon has been calculated using the time-dependent configuration-interaction-singles method and was compared to the random-phase-approximation method \cite{Karamatskou/PRA:2017}. 
In a recent work \cite{Hofbrucker/PRA:2016}, moreover, we investigated the two-photon ionization of neutral atoms and demonstrated, that the screening of the active electron by all other electrons can significantly alter the total cross section of light elements such as O, F, Ne, Na, or Mg. This decrease is a direct consequence of a drop of the dominant ionization channel. As shown below, the behavior of the dominant channel has an even stronger impact upon the photoelectron angular distributions. In Ref. \cite{Hofbrucker/NIMB:2017}, it was shown that the relativistic wavefunction contraction makes the strongest (relativistic) effect in calculating the total two-photon ionization cross sections.

In this contribution, we investigate the photoelectron angular distributions of the non-resonant two-photon $K$-shell ionization of neutral atoms. In Sec. \ref{Sec.Theory}, we describe our theoretical approach and provide a first intuitive view of the problem. In Sec. \ref{Sec.ResultsAndDiscussion}, we first present typical photoelectron angular distributions for ionization of atoms by circularly, linearly, elliptically, and unpolarized light. Then, we demonstrate that significant deviations from the typical distributions can arise due to the screening or relativistic effects. We show, in particular, that the screening effects can lead to an \textit{elliptical} dichroism, and discuss the possibility of performing a complete experiment which would test this prediction. Finally, a summary is given in Sec. \ref{Sec.SummaryAndOutlook}.

Relativistic units ($\hbar=c=m=1$) are used throughout the paper, unless stated otherwise.

\section{Theoretical background}

We study here the process of the non-resonant two-photon $K$-shell ionization of neutral atoms.
For the sake of simplicity, we shall start directly from a \textit{single-active}-electron representation, cf. see \cite{Hofbrucker/PRA:2016, Hofbrucker/NIMB:2017} for further details. In particular, we here derive a relativistic angle-differential cross section in which the geometry (of possible observations) is clearly separated from the physical interaction. In Sec. \ref{SubSec.ApproximateParametrization}, this expression is simplified by applying the nonrelativistic limit and electric dipole approximation in order to obtain a simple formula for the photoelectron angular distribution, suitable also for parametrization. In Sec. \ref{SubSec.Computation}, we present the computational approach which was used to obtain our results.

\label{Sec.Theory}
\subsection{General theory}

We employ the independent particle approximation, in which we assume that only one of the $K$-shell electrons interacts with the two photons. This electron-photon interaction is treated within the second-order perturbation theory, while the effect of all other (inactive) electrons of the neutral atom are accounted for by a screening potential included in the Hamiltonian of the Dirac equation. In this single-active-electron representation, the two-photon ionization process can be represented as follows
\begin{eqnarray}\label{Eq.Process}
\ketm{n_a \kappa_a m_a}+\gamma(\bm{k}_1, \lambda_1)+\gamma(\bm{k}_2, \lambda_2) \rightarrow \ketm{\bm{p}_e m_e}.
\end{eqnarray}
Since we consider the ionization of a $K$-shell electron, the principle quantum number $n_a$, Dirac quantum number $\kappa_a$ and the projection of the total angular momentum $m_a$ of the initial bound electron take the values $n_a=1$, $\kappa_a=-1$, and $m_a=\pm\frac{1}{2}$. The Dirac quantum number $\kappa$ is defined by the total and orbital angular momenta $j$ and $l$ as $\kappa=\mp(j+\frac{1}{2})$ for $j=l\pm \frac{1}{2}$. In a typical experiment, the two photons in process (\ref{Eq.Process}) originate from the same source, and have equal wavevectors $\bm{k}_1=\bm{k}_2=\bm{k}$. The polarization of the photons is conveniently described in the helicity ($\lambda$) representation, where it can be fully characterized by its density matrix $\mem{\bm{k} \lambda}{\hat{\rho}}{\bm{k} \lambda'}$ in terms of the linear $P_1, P_2$, and circular $P_3$ Stokes parameters \cite{Blum/Book:1981}. Due the interaction of the initially bound electron with the two photons, the electron is promoted into a continuum state $\ketm{\bm{p}_e m_e}$, with a well-defined asymptotic momentum $\bm{p}_e$, and spin projection $m_e$. 
In lowest order perturbation theory (and within the independent particle approximation), the transition element for the interaction of the neutral atom with two photons can be written as
\begin{eqnarray}\label{GeneralTransitionAmplitude}
M_{m_e m_a}^{\lambda_1 \lambda_2}=\int\kern-1.5em\sum_{~n}\frac{
	\mem{\bm{p}_e m_e}
		{\bm{\alpha \cdot A}_{\lambda_2}}
		{n}
	\mem{n}
		{\bm{\alpha \cdot A}_{\lambda_1}}
		{n_{a} \kappa_a m_a}		
		}
	{E_{n_a \kappa_a}+\omega-E_{n_n \kappa_n}}, \nonumber\\
\end{eqnarray}
where $\bm{\alpha \cdot A}_{\lambda}$ represents the electron-photon interaction operator, $\omega$ is the photon energy, and a summation over the complete spectrum of intermediate states $\ketm{n}\equiv~\ketm{n_{n}\kappa_{n}m_{n}}$ needs to be carried out. In order to calculate the cross section, it is necessary, moreover, to carry out the multipole expansion of the free electron wavefunction and the electron-photon operator, see Ref. \cite{Hofbrucker/NIMB:2017} for details. Making use of these expansions, we obtain
\begin{eqnarray}\label{Eq.MatElExpanded}
M^{\lambda_1 \lambda_2}_{m_e m_a}&=&
4\pi \sum_{\substack{J_1 p_1\\ J_2 p_2}}\sum_{q m_q} \sum_{\kappa m_j m_l} \sum_{\kappa_n}  
(-1)^{j+q-m_j}i^{J_1+J_2-l}e^{i\Delta_\kappa}
\nonumber \\ &\times&  \sprm{j_n 1/2 J_2 0}{j 1/2} \sprm{j_a 1/2 J_1 0}{j_n 1/2}  Y_{l m_l}(\bm{\hat{p}}_e)
\nonumber \\ &\times& 
 \sprm{l m_l 1/2 m_e}{j m_j} \sprm{j -m_j j_a m_a}{q m_q} 
\nonumber \\ &\times& 
 [J_1,J_2,jn,j_a]^{1/2} \Pi_{l_a,J_1,l_n, p_1} \Pi_{l_n, J_2, l, p_2}
 \nonumber \\ &\times& 
 \sixjm{j}{j_a}{q}{J_1}{J_2}{j_n} T_{qm_q}^{p_1 p_2}(\lambda_1, \lambda_2) U_{\kappa_a \kappa_n \kappa}, 
\end{eqnarray}
where the typical notations $\sixjm{.}{.}{.}{.}{.}{.}$ and $\sprm{....}{..}$ are used to represent the six-j symbols and Clebsch-Gordan coefficients, $[x_1, x_2, ...]=(2x_1+1)(2x_2+1)....$, $J'$s are the multipoles of the two photons, and the index $p$ describes the electric ($p=1$) or magnetic ($p=0$) component of the electromagnetic field. Furthermore, $l$ and $j$ arise from the partial-wave expansion of the free electron wavefunction and represent its orbital and total angular momenta, $\Delta_{\kappa}$ is the corresponding phase factor \cite{Eichler/PR:2007}, $Y_{lm}$ are the spherical harmonics, and $U_{\kappa_a \kappa_n \kappa}$ represents the radial part of the transition amplitude (\ref{GeneralTransitionAmplitude}) for a specific ionization channel $\kappa_a \rightarrow \kappa_n \rightarrow \kappa$, see e.g. Eqs. 6.129 in Ref. \cite{Johnson/Book} for an explicit expression of the radial integrals. We also introduced an intermediate angular momentum $q$, which represents the transfer of angular momenta between the initial $j_a$ and final $j$ states. The momentum $q$ then also corresponds to the momentum transfer from the two photons with multipoles $J_1$ and $J_2$ to the electron. The functions $\Pi$ are defined as $\Pi_{l_1,l_2,l_3,p}=1$ if the sum $l_1+l_2+l_3+p$ is odd, and $\Pi_{l_1,l_2,l_3,p}=0$ otherwise. Finally, the irreducible tensors $T_{q}^{p_1 p_2}(\lambda_1, \lambda_2)$ are defined similarly as in Refs. \cite{Manakov/JPB:2002, Surzhykov/PRA:2011}
\begin{eqnarray}
T_{q}^{p_1, p_2}(\lambda_1,  \lambda_2)=(-i)^{p_1+p_2}\{ [\bm{\hat{\varepsilon}}_{\lambda_1}\cdot\bm{Y}_{J_1}^{(p_1)}]\otimes \nonumber [\bm{\hat{\varepsilon}}_{\lambda_2}\cdot\bm{Y}_{J_2}^{(p_2)}]\}_{q}, \\
\end{eqnarray}
with $\bm{\hat{\varepsilon}}_{\lambda}$ being the unit polarization vectors and
$\bm{Y}^{(p)}_{J}$ representing the vector composed by rank-$J$ tensors of spherical harmonics. 
The angle-differential cross section for two-photon ionization of an unpolarized atom is then given by
\begin{eqnarray}\label{Eq.CrossSection}
\frac{d\sigma}{d\Omega}&=&
\frac{8 \pi^3 \alpha^2}{\omega^2}\frac{1}{[j_a]}\sum_{\substack{\lambda_1\lambda_2\\\lambda'_1\lambda'_2}}
\mem{\bm{k} \lambda_1}{\hat{\rho}}{\bm{k} \lambda'_1}
\mem{\bm{k} \lambda_2}{\hat{\rho}}{\bm{k} \lambda'_2}\nonumber \\
&\times & \sum_{m_e m_a} M_{m_e m_a}^{\lambda_1 \lambda_2} M_{m_e m_a}^{\lambda'_1 \lambda'_2 *}. 
\end{eqnarray}
After coupling of the angular momenta $q$ and $q'$ of the two transition amplitudes of the above equation with the momentum $L$, and further tensor manipulation \cite{Varshalovich/Book:1988, Manakov/JPB:2002}, it is possible to separate the geometrical and the structural characteristics of the cross section. For the sake of simplicity, we introduce the notation where the dependence on q implies the dependence on $J_1$ and $J_2$. Then we can write the cross section as
\begin{eqnarray}\label{Eq.CrossSectionSeparated}
\frac{d\sigma}{d\Omega}=
\frac{8 \pi^3 \alpha^2}{\omega^2}\frac{1}{[j_a]}
\sum_{L}
\sum_{\substack{p_1 p'_1\\ p_2 p'_2}} 
\sum_{\substack{\{J_1 J_2\}q\\ \{J'_1 J'_2\}q'}} 
\mathfrak{G}^{p_1 p_1' p_2 p'_2}_{q q';L} \mathfrak{F}_{q q'; L }^{p_1 p'_1 p_2 p'_2}.~~~
\end{eqnarray}
Here, the $\mathfrak{G}_{L}$ represents the "geometrical part" of the cross section, which completely defines the photon polarization, as well as all angular characteristics of the two-photon ionization process, i.e. it contains both, the photon and the photoelectron angular dependencies. The explicit form of this term is given by
~\\
~\\
\begin{eqnarray}\label{Eq.G}
\mathfrak{G}^{p_1 p_1' p_2 p'_2}_{q q'; L}&=& (-1)^{p'_1+p'_2}
\sum_{\substack{\lambda_1\lambda_2\\ \lambda'_1\lambda'_2}}
\mem{\bm{k} \lambda_1}{\hat{\rho}}{\bm{k} \lambda'_1}
\mem{\bm{k} \lambda_2}{\hat{\rho}}{\bm{k} \lambda'_2} \nonumber\\ &\times&
[\{T_{q}^{p_1 p_2}(\lambda_1,\lambda_2)\otimes T_{q'}^{p'_1 p'_2}(\lambda'_1,\lambda'_2)\}_L \cdot Y_L(\bm{\hat{p}}_e)]. \nonumber\\ 
\end{eqnarray}
It is worth noting, that up to now, no choice of geometry has been made, hence the geometrical part is generally applicable for any choice of the quantization axis. The "structural part" $\mathfrak{F}_L$ of the cross section encapsulates the properties of the electron-photon interaction, and depends on the details of the (radial) wave functions. It is given by \vspace{0.8cm}

\begin{widetext}

\begin{eqnarray}\label{Eq.F}
\mathfrak{F}_{q q'; L}^{p_1 p'_1 p_2 p'_2}=
\sum_{\kappa  \kappa'} (4\pi)^{3/2} (-1)^{j_a+1/2+L+q} [L]^{-1/2}
\sprm{l0l'0}{L0} \sixjm{l}{L}{l'}{j'}{1/2}{j}\sixjm{q}{L}{q'}{j'}{j_a}{j} 
R_{\kappa_a \kappa; q}^{} R_{\kappa_a \kappa'; q'}^{*},
\end{eqnarray}
where 
\begin{eqnarray}\label{Eq.R}
R_{\kappa_a \kappa; q}&=&\sum_{\kappa_n} (-1)^{j}[j_a,l,q,j,j_n,J_1,J_2]^{1/2}i^{J_1+J_2+l}e^{i\Delta_{\kappa}}\Pi_{l_a,J_1,l_n, p_1} \Pi_{l_n, J_2, l, p_2} \nonumber\\ 
&\times& \sprm{j_n 1/2 J_2 0}{j 1/2}\sprm{j_a 1/2 J_1 0}{j_n 1/2} \sixjm{j}{j_a}{q}{J_1}{J_2}{j_n} U_{\kappa_a \kappa_n \kappa}.
\end{eqnarray}
\end{widetext}
By looking at the expression above as well as Eq. (\ref{Eq.MatElExpanded}), and using the properties of Clebsch-Gordan coefficients, we deduce the restrictions on the coupling parameters $q=0,2$ and $L=0, 2, 4$.

\subsection{Non-relativistic limit of the differential cross section}
\label{SubSec.ApproximateParametrization}

Expressions (\ref{Eq.CrossSectionSeparated}-\ref{Eq.R}) describe the angle-differential cross section of the two-photon ionization, and are general within the framework of the independent-particle model. In the derivation of these expressions, no assumptions were made with regard to the choice of a quantization axis nor photon polarization, but for the price of obtaining a rather complex expression. For many atoms and ions, however, a nonrelativistic description is completely sufficient \cite{Manakov/JETP:2009}, i.e. for systems where the relativistic effects are negligible or small . In Ref. \cite{Manakov/JETP:2009}, the two-photon above threshold ionization was studied within nonrelativistic theory, and a parametrized expression for the angle-differential cross section was derived. However, this cross section expression is valid only for ionization of atoms by fully polarized photons. In this section, we will show that a similar expression can be obtained for a general photon polarization by applying a number of reasonable assumptions to the cross section from Eq. (\ref{Eq.CrossSectionSeparated}). 

\begin{figure}[t]
\includegraphics[scale=0.50]{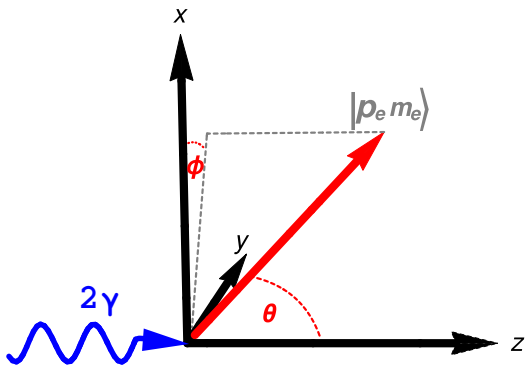}
\caption{Geometry of the considered two-photon ionization process within the considered geometry. An initially $K$-shell bound electron of a neutral atom absorbs two photons, and is subsequently promoted into the continuum. The quantization $z$-axis is defined by the propagation direction of the photons and $x$-axis is chosen to be alligned with the linear component of the photon polarization. For the ionization by circular or unpolarized light, the cross section does not depend on the choice of $x$-axis. The photoelectron is emitted into a direction given by the polar and azimuthal angles $\theta$ and $\phi$, respectively.}\label{Fig.ProcessDiagram}
\end{figure}

The general cross section expression can be simplified if we choose some appropriate geometry. For example, we take the quantization $z$-axis along the photon propagation direction $\bm{\hat{k}}$ and choose the linear component of the photon polarization to be aligned with the $x$-axis, then we characterize the propagation direction of the photoelectron $\bm{\hat{p}}_e$ by the polar and azimuthal angles $\theta$ and $\phi$, respectively, see Fig.~\ref{Fig.ProcessDiagram}. Furthermore, we shall restrict ourselves to the electric-dipole approximation, i.e. taking $J_1=J_2=p_1=p_2=1$ in Eq. (\ref{Eq.CrossSectionSeparated}). As we are considering the two-photon ionization of an $s$-state electron, only two possible ionization paths are possible within this approximation; $s \rightarrow p \rightarrow s$ and $s \rightarrow p \rightarrow d$. In a relativistic framework, in contrast, the fine structure splits these two paths into five paths. In the nonrelativistic limit, however, the transition amplitudes as well as phase factors remain unaffected by the fine structure splitting, thus the five relativistic paths $U_{\kappa_a \kappa_n \kappa}$ reduce to two nonrelativistic ones; $U_s$ and $U_d$. Since phase factors depend on angular momenta of the final photoelectron state only, the three relativistic phases $\Delta_\kappa$ reduce to $\Delta_s$ and $\Delta_d$. With these assumptions in mind and by performing all summations in Eq. (\ref{Eq.CrossSectionSeparated}), we obtain 

\begin{figure*}[t]

\minipage{0.25\textwidth}
 \includegraphics[width=1.0\textwidth]{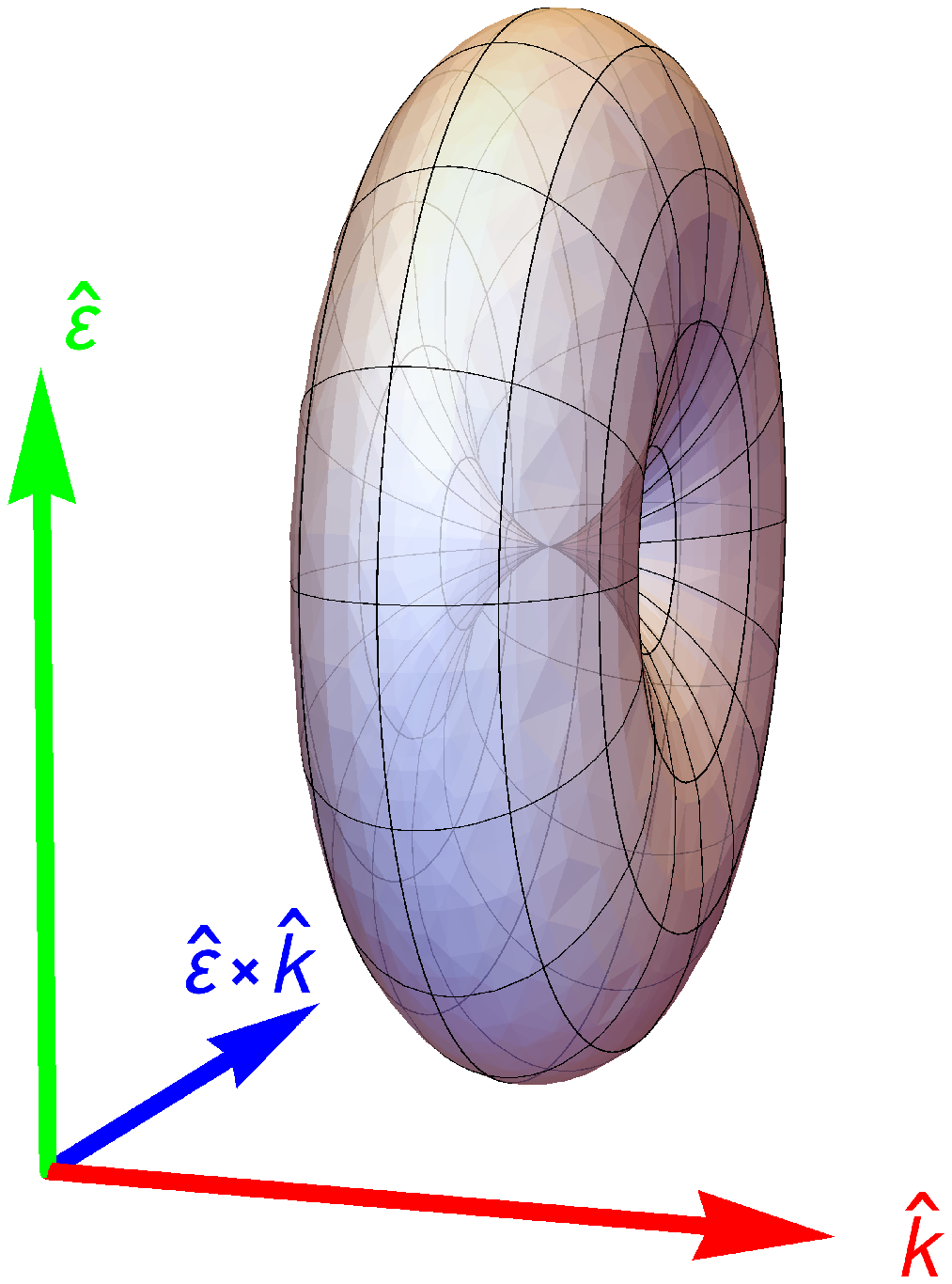}
\endminipage\hfill
\minipage{0.25\textwidth}
  \includegraphics[width=1.0\textwidth]{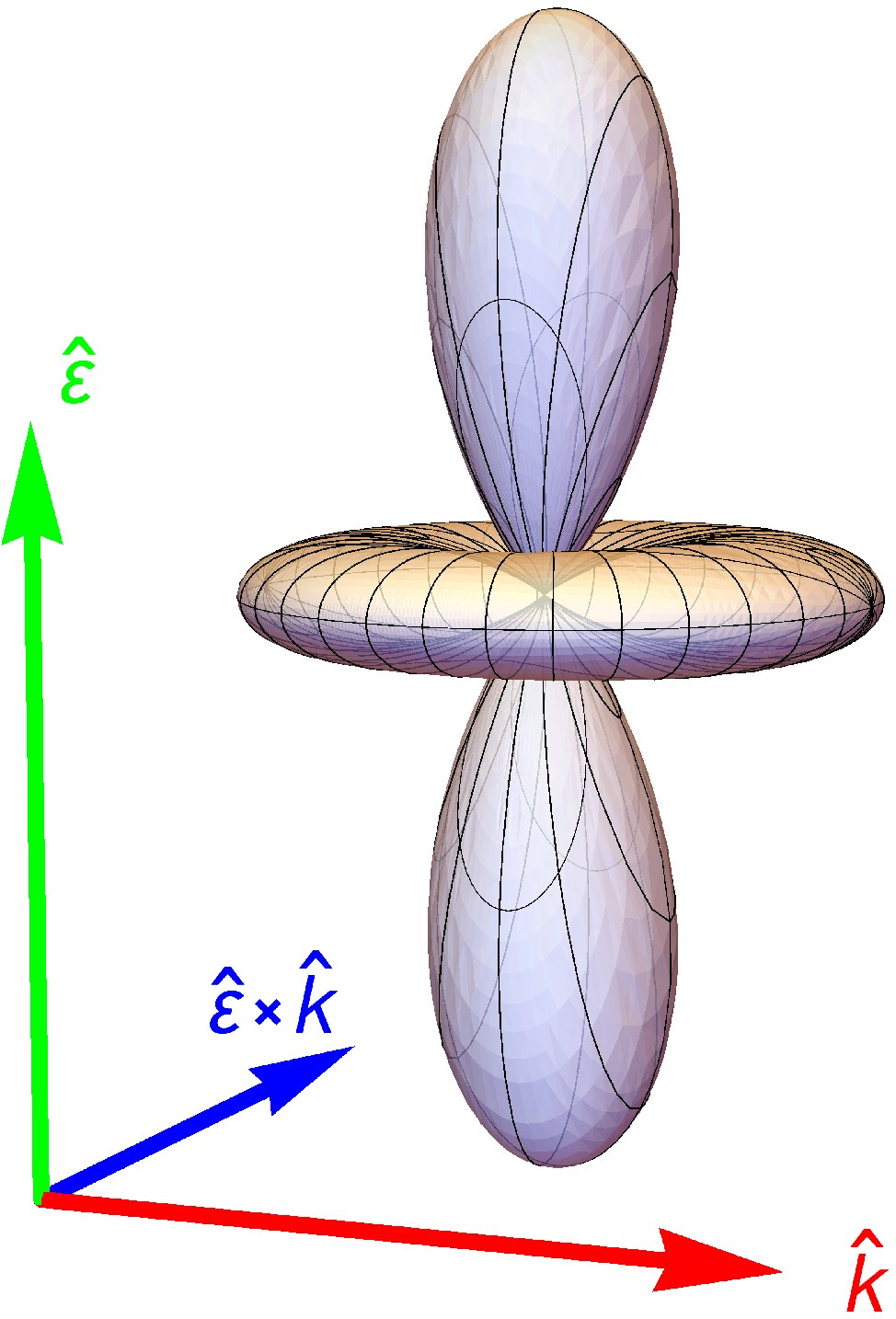}
\endminipage\hfill
\minipage{0.25\textwidth}
  \includegraphics[width=1\linewidth]{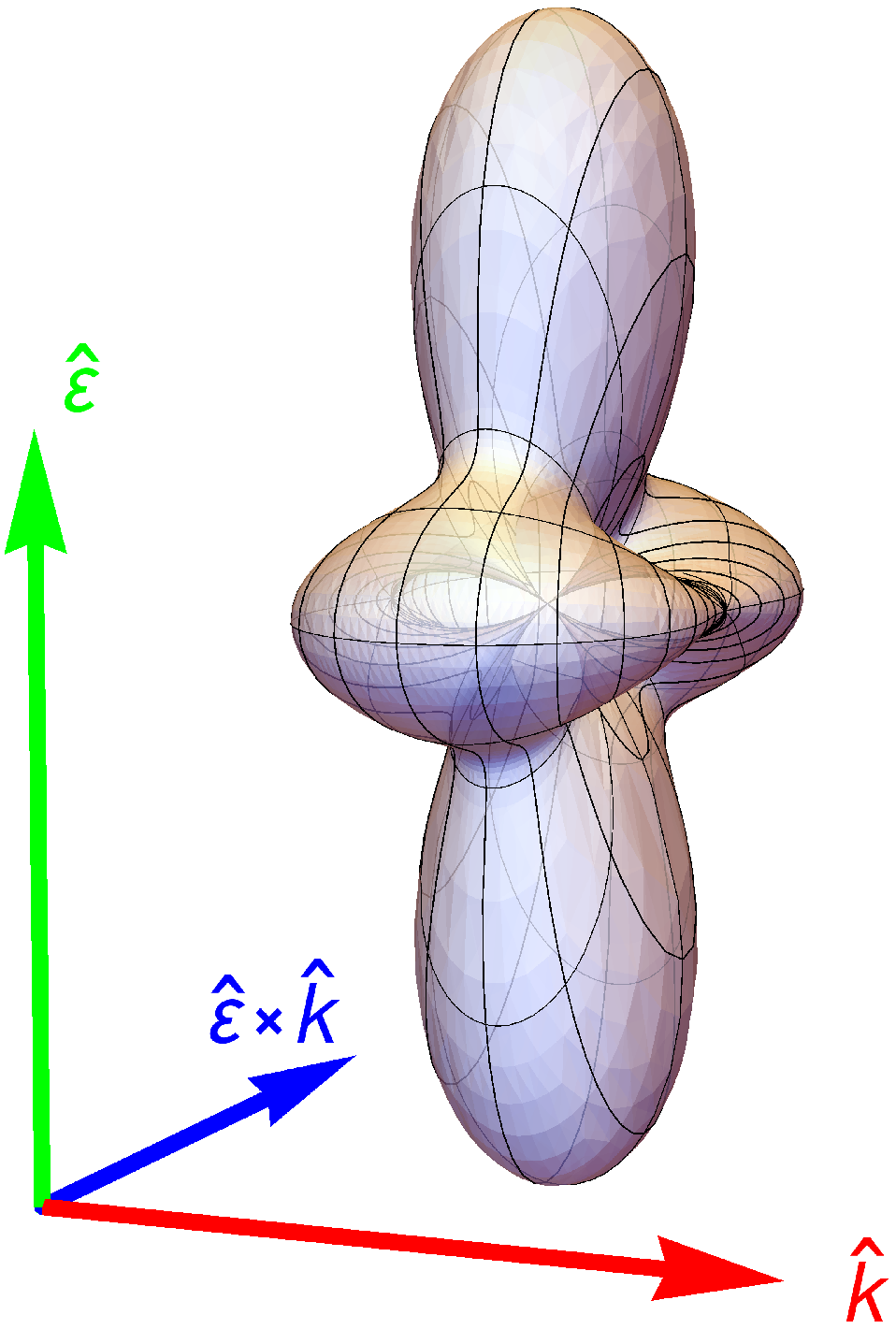}
\endminipage\hfill
\minipage{0.25\textwidth}
  \includegraphics[width=1\linewidth]{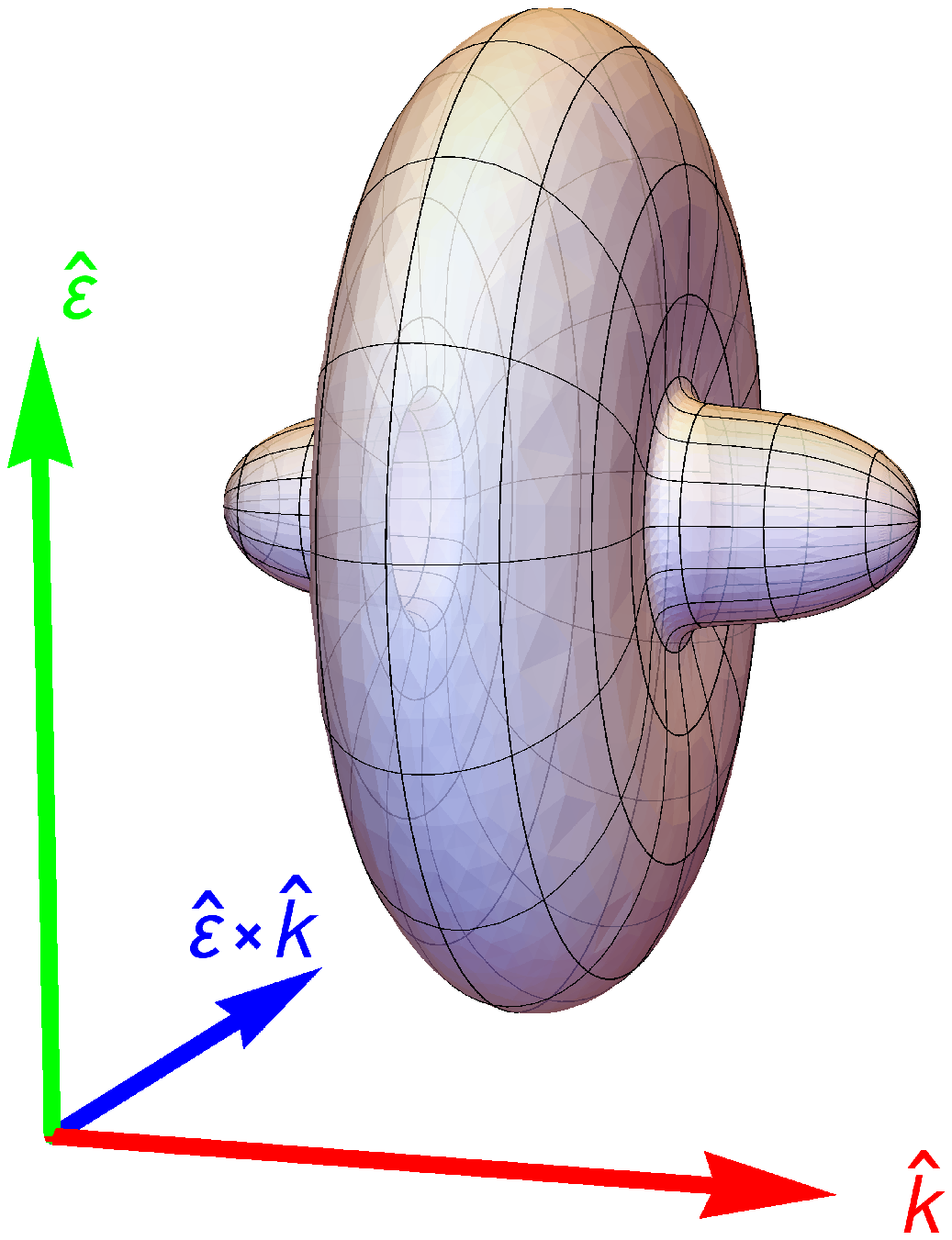}
\endminipage\hfill

\caption{Typical photoelectron angular distributions for the nonresonant two-photon $K$-shell ionization of neutral atoms for incident photons polarized circularly ($P_c=1$), linearly($P_l=1$), elliptically ($P_c=1/\sqrt{2}$, $P_l=1/\sqrt{2}$), or unpolarized ($P_c=0$, $P_l=0$).}\label{Fig.3d}
\end{figure*}

\begin{eqnarray}\label{Eq.E1E1}
\frac{d\sigma^{(nonrel.)}}{d\Omega}&=&
\frac{9 \pi^2 \alpha^2}{2 \omega^2} \Big \{
|U_s|^2 \mathcal{P}+|U_d|^2\Big[ \mathcal{P}-3\mathrm{sin}^2\theta \big(\mathcal{P}+\nonumber\\
&+&2P_l\mathrm{cos}(2\phi)\big)+\frac{9}{2}\mathrm{sin}^4\theta\big(1+P_l\mathrm{cos}(2\phi)\big)^2 \Big]+\nonumber \\
&+&2\mathrm{Re}\big[U_s U_d^*e^{i(\Delta_s-\Delta_d)}[ \mathcal{P}-\frac{3}{2}\mathrm{sin}^2\theta \big(\mathcal{P} + \nonumber \\
&+&2P_l\mathrm{cos}(2\phi)+2iP_l P_c \mathrm{sin}(2\phi)\big)]\Big]\Big \},
\end{eqnarray}
where the degree of linear polarization $P_l$ can be represented in terms of Stokes parameters as $P_l=\sqrt{P_1^2+P_2^2}$, the degree of circular polarization is $P_c=P_3$, and $\mathcal{P}=1+P_l^2-P_c^2$. For the case of completely polarized photons, i.e. $P_l^2+P_c^2=1$, the above formula reduces to the one derived in Ref. \cite{Manakov/JETP:2009}. The indexes of the radial integrals $U_s$ and $U_d$ as well as the corresponding nonrelativistic phases $\Delta_s$ and $\Delta_d$ refer to the corresponding photoelectron $s-$ and $d-$partial waves.

 If we integrate Eq. (\ref{Eq.E1E1}) over the angles $\theta$ and $\phi$, we obtain the total cross section for ionization of atoms by two-photons of general polarization
\begin{eqnarray}
\sigma^{(nonrel.)}=\frac{18 \pi^3 \alpha^2}{5\omega^2}(5\mathcal{P}|U_s|^2  + (7+P_l^2+5P_c^2)|U_d|^2). \nonumber \\
\end{eqnarray}
This compact expression depends solely on the two radial matrix elements $U_s$ and $U_d$. In the case of fully polarized photons, this equation reduces to the well-known expressions of Ref. \cite{Arnous/PRA:1973}. It can be seen that in order to obtain the information about the photoelectron phase, one needs to measure the corresponding angular distribution, as the total cross section does not depend on the phase.

\subsection{Computation}
\label{SubSec.Computation}
The evaluation of the cross section (\ref{Eq.CrossSectionSeparated}) requires an infinite summation over all multipole orders of the electron-photon interaction operator as well as over the complete energy spectrum of intermediate states. The infinite summation over higher multipoles converges at the fifth order, therefore summation over more terms was not necessary. To sum over the infinite number of intermediate states, finite basis-set \cite{Sapirstein/JPB:1996} constructed from $B$-splines by applying  the dual-kinetic-balance approach \cite{Shabaev/PRL:2004} was employed. This technique allows us to reduce the infinite summation over the intermediate states to a finite sum over a pseudospectrum. This approach has been previously successfully applied, for example, in the calculations of two-photon decay processes of heliumlike ions \cite{Volotka/PRA:2011, Volotka/PRL:2016} or cross sections of x-ray Rayleigh scattering \cite{Volotka/PRA:2016}. The continuum-state wavefunctions were obtained by numerical solutions of the Dirac equation with help of the RADIAL package \cite{Salvat/CPC:1995}. In order to account for the screening effects, we solve the Dirac equation with a screening potential, which partially accounts for the interelectronic interaction. We use the core-Hartree potential, which corresponds to a potential created by all bound electrons except of the active electron. In Ref. \cite{Hofbrucker/PRA:2016}, it has been shown that the choice of a screening potential does not significantly affect the total cross section of the observed behavior of the process. Our current results confirm, that the same statement holds true for the photoelectron angular distribution. The screening potential accounts for the major part of the many-electron contributions, while contributions beyond the independent particle approximation are expected to be negligible, similarly as for the case of Rayleigh scattering \cite{Volotka/PRA:2016}.

%
%
%
\section{Results and discussion}
\label{Sec.ResultsAndDiscussion}

In the two-photon $K$-shell ionization, the photoelectron angular distribution shows more often than not the same behavior, quite independent of the atomic target and the coupling of the valence-shell electrons. Figure \ref{Fig.3d} displays such typical distributions for four different photon polarizations; circular, linear, elliptical, as well as for unpolarized photons. For the ionization by circularly or unpolarized photons, obviously, the photoelectron angular distributions are always axially symmetric and, thus, independent of the azimuthal angle $\phi$. We also note, that there is no photoelectron emission along the photon propagation direction for the ionization of unpolarized atoms by two completely circularly polarized photons. Indeed, the emission along this axis is forbidden by the conservation of projection of angular momentum. Since the helicity of the two photons is $\lambda_1+\lambda_2=\pm 2$, the change in the projection of angular momentum cannot be compensated by the photoelectron emitted along the photon propagation direction. However, for ionization by photons with a lower degree of circular polarization ($P_c<1$), the electron emission along the quantization axis becomes possible, and the distribution will become similar to the one for the unpolarized case. This can be also seen analytically from Eq. (\ref{Eq.E1E1}). For the completely circularly polarized case ($P_c=1$), we have $\mathcal{P}=0$ and $P_l=0$. Therefore, the photoelectron distribution is given solely by the sin$^4 \theta$ distribution, which corresponds to $l=2,~m=\pm 2$ partial wave of the photoelectron. If we decrease the polarization purity, however, other partial waves will also contribute to the photoelectron distribution and the emission into forward direction will increase. For the case of two-photon ionization of atoms by linearly polarized light, the photoelectrons are dominantly emitted along the photon polarization direction. From Eq. (\ref{Eq.E1E1}), we see that  the distribution now depends also on the azimuthal angle $\phi$, and that it contains contributions from both ionization channels. 

Although the distributions of Fig. \ref{Fig.3d} generally provide a good description of the photoelectron emission direction, we will present cases, where significant deviations from these distributions occur due to screening or relativistic effects. While the screening effects are taken into account also in the nonrelativistic cross section (\ref{Eq.E1E1}), which can be characterized by two parameters, the expression is insufficient to describe relativistic processes. It is the aim of this section to critically evaluate the validity of this nonrelativistic description and show its limitations. Below, we shall assess also the importance of the screening potential and compare these exact calculation with a calculation, where no account for the inactive electrons was made. 
Although the formulas (\ref{Eq.CrossSectionSeparated}) and (\ref{Eq.E1E1}) are generally applicable for any two-photon ionization process, detailed calculations have been carried out just for the two-photon $K$-shell ionization of neutral neon, germanium, xenon, and uranium atoms, and for photon energies below the 1$s$-2$p$ resonance. In order to compare the results for different elements, we shall present our data in terms of excess energy $\varepsilon$, i.e. the ratio of the energies of the (two) incident photons and the $K$-shell binding energy ($E_b$), $\varepsilon=2 \omega/E_b$.

\begin{figure*}
\includegraphics[width=\linewidth]{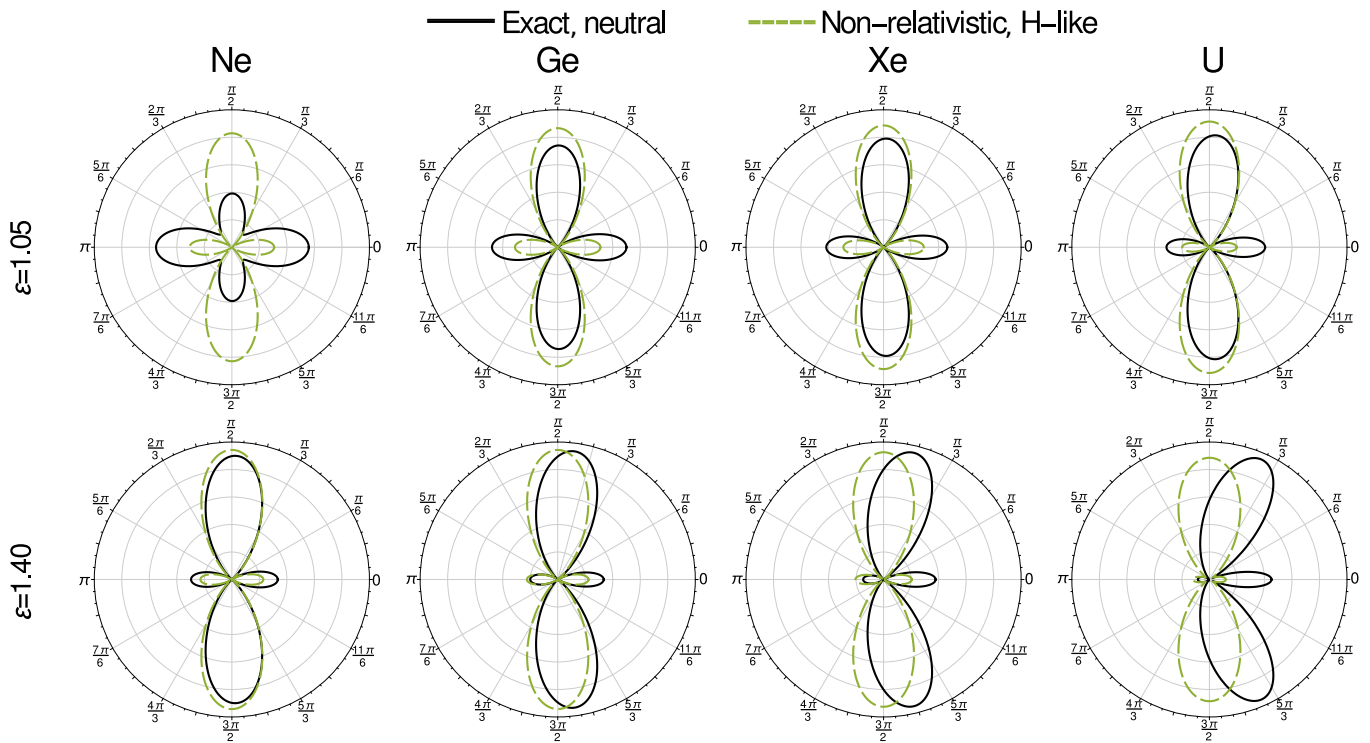}
\caption{Screening and multipole contributions to the two-photon $K$-shell ionization of neutral Ne, Ge, Xe, and U atoms, using linearly polarized photons with $\varepsilon=1.05$ (upper row) and $\varepsilon=1.40$ (lower row) excess energies. An exact relativistic computation of two-photon ionization of neutral atoms (solid black) is compared with a nonrelativistic calculation for ionization of H-like ions (dashed green) is presented. At $\varepsilon=1.05$ excess energy, the photoelectron emission from a neon atom (upper-left figure) significantly decreases along the polarization axis due to the screening effects.  
For uranium $\varepsilon=1.40$ (lower-right figure), the emission into the incident photon direction is highly promoted due to relativistic effects. In order to compare the results for neutral atoms and H-like ions, the distributions have been scaled in the given plane. All plots were obtained for $\phi=0$, i.e. in the $\bm{\hat{\varepsilon}}\bm{\hat{k}}$-plane.}\label{Fig.Effects}
\end{figure*}

\subsection{Relativistic and screening effects}

Figure \ref{Fig.Effects} presents the photoelectron angular distributions of the two-photon $K$-shell ionization of neutral and H-like atoms for two excess energies ($\varepsilon=1.05,~1.40$) and for four elements (Ne, Ge, Xe, U). The solid black figures correspond to the relativistic calculations of two-photon ionization of neutral atoms, while the dashed green figures correspond to the nonrelativistic calculations of two-photon ionization of H-like ions. The results of the latter serve as reference distributions, since they are neither affected by the relativistic nor screening effects. Therefore, the comparison of the corresponding results for these two calculations gives us the necessary insight how the relativistic and multipole contributions affect the angular emission of photoelectrons in the two-photon $K$-shell ionization of atoms and ions.

There are various relativistic contributions to the cross sections. For the total cross section, the relativistic contraction of the wavefunction may result in a reduction of up to 30$\%$, while the higher multipoles give rise to rather small changes only \cite{Hofbrucker/NIMB:2017}. For the photoelectron angular emission, in contrast, the different multipole contributions may significantly alter the distribution. These contributions sensitively depend on the nuclear charge and the energy of the incident photons. For ionization of medium and heavy atoms with high energetic photons, a forward emission of the photoelectron is enhanced, and the backward emission decreases.
This distortion of the angular distribution can be clearly seen in Fig. \ref{Fig.Effects}, together with the nuclear charge and photon energy dependencies. Although, we present results for incident linearly polarized light, an identical change of the distributions is found for all types of polarization. A similar behavior was found in relativistic calculations of H-like atoms \cite{Koval/JPB:2004}. Interestingly, the distortion of the distribution due to the multipole effects in two-photon ionization of neutral atoms is comparable with the one for H-like ions. This can be understood from Fig. \ref{Fig.Effects}. Since the importance of multipole contributions depends more strongly on the photon energy, the results do not significantly differ if the nuclear potential is partially screened by the electrons in higher shells.

\begin{figure*}[!htb]
\minipage{0.3\textwidth}
  \includegraphics[width=\linewidth]{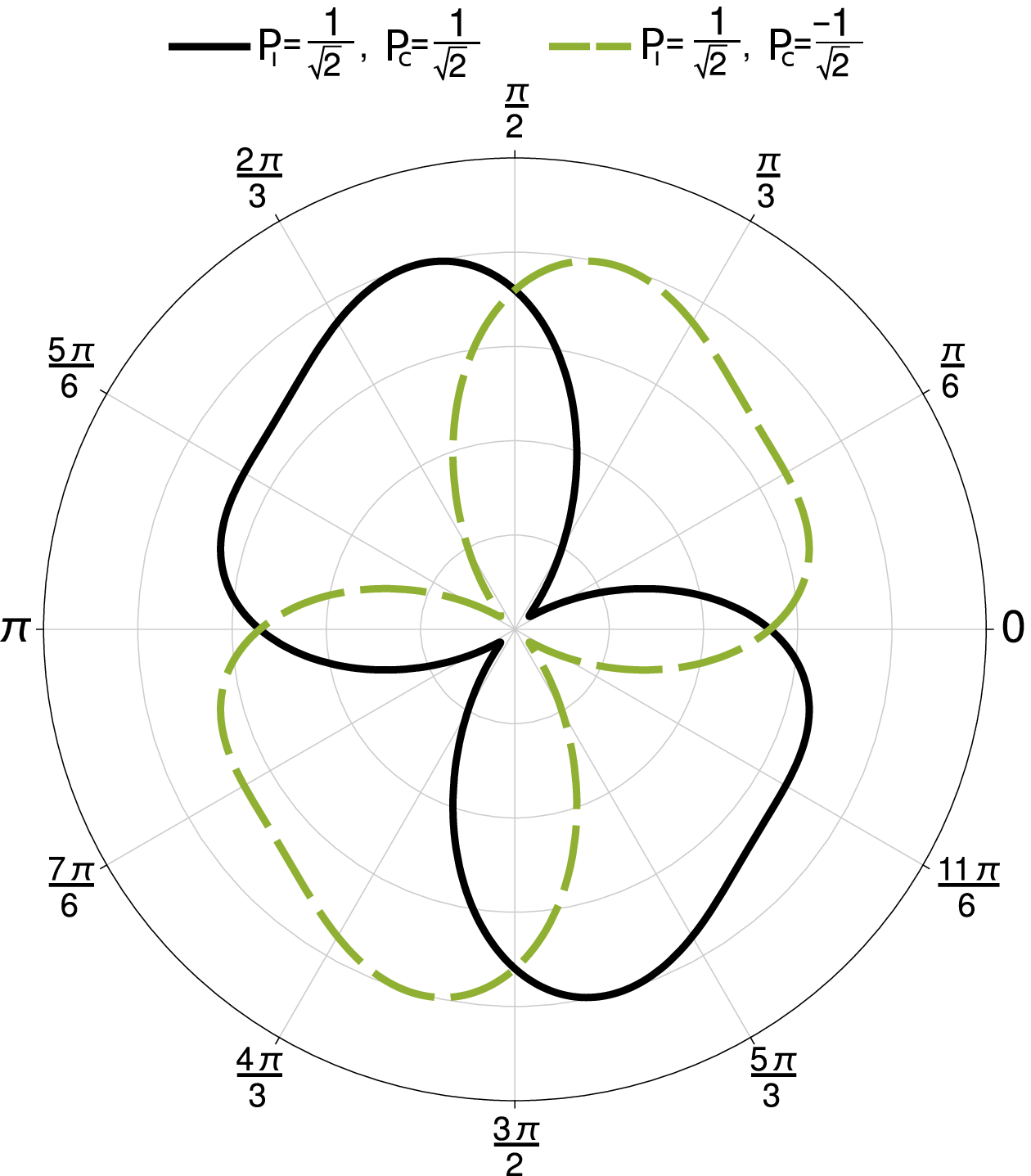}
\endminipage\hfill
\minipage{0.32\textwidth}

  \includegraphics[width=\linewidth]{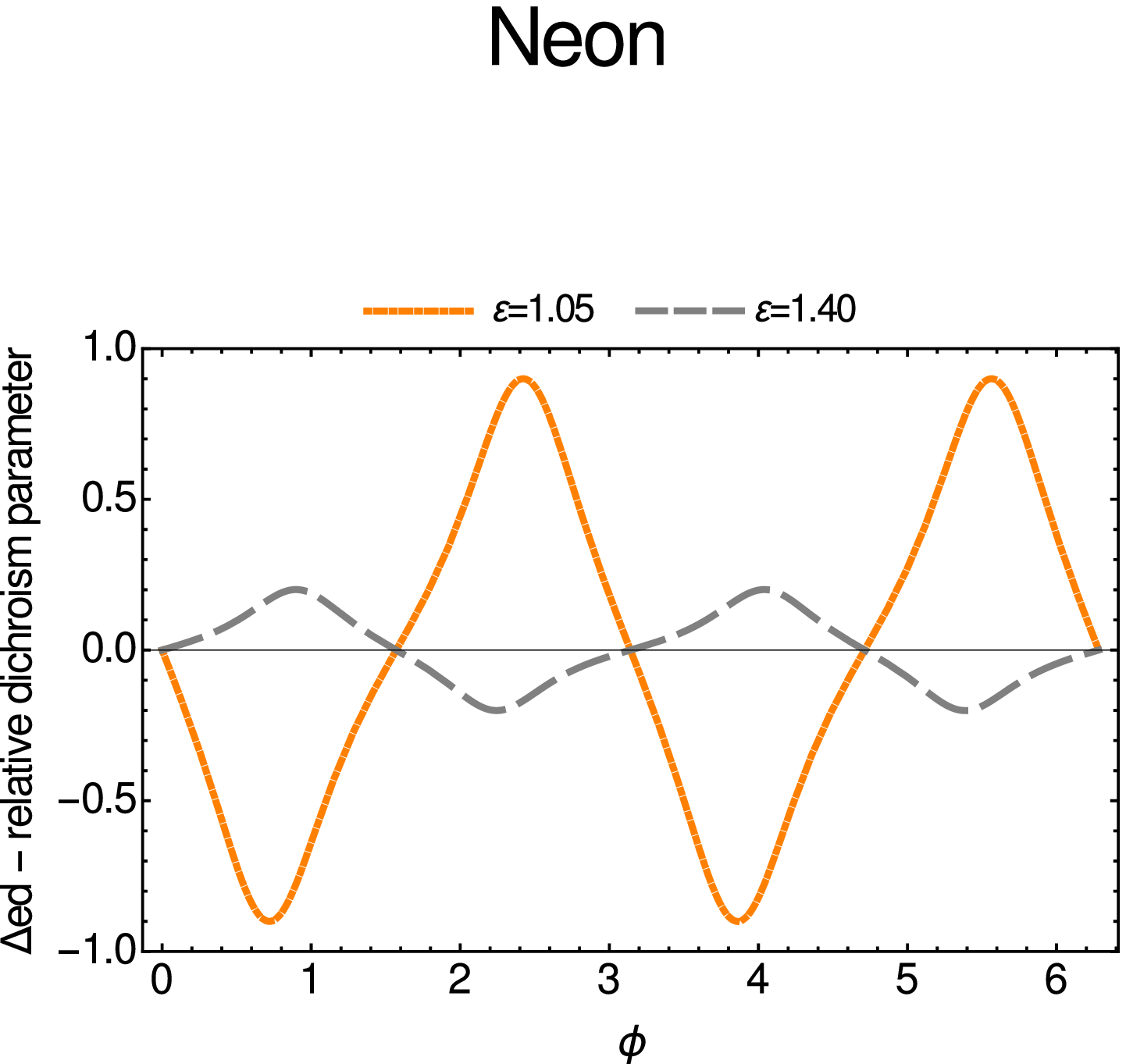}
  \vspace{0.35cm}
\endminipage\hfill
\minipage{0.3\textwidth}%
  \includegraphics[width=\linewidth]{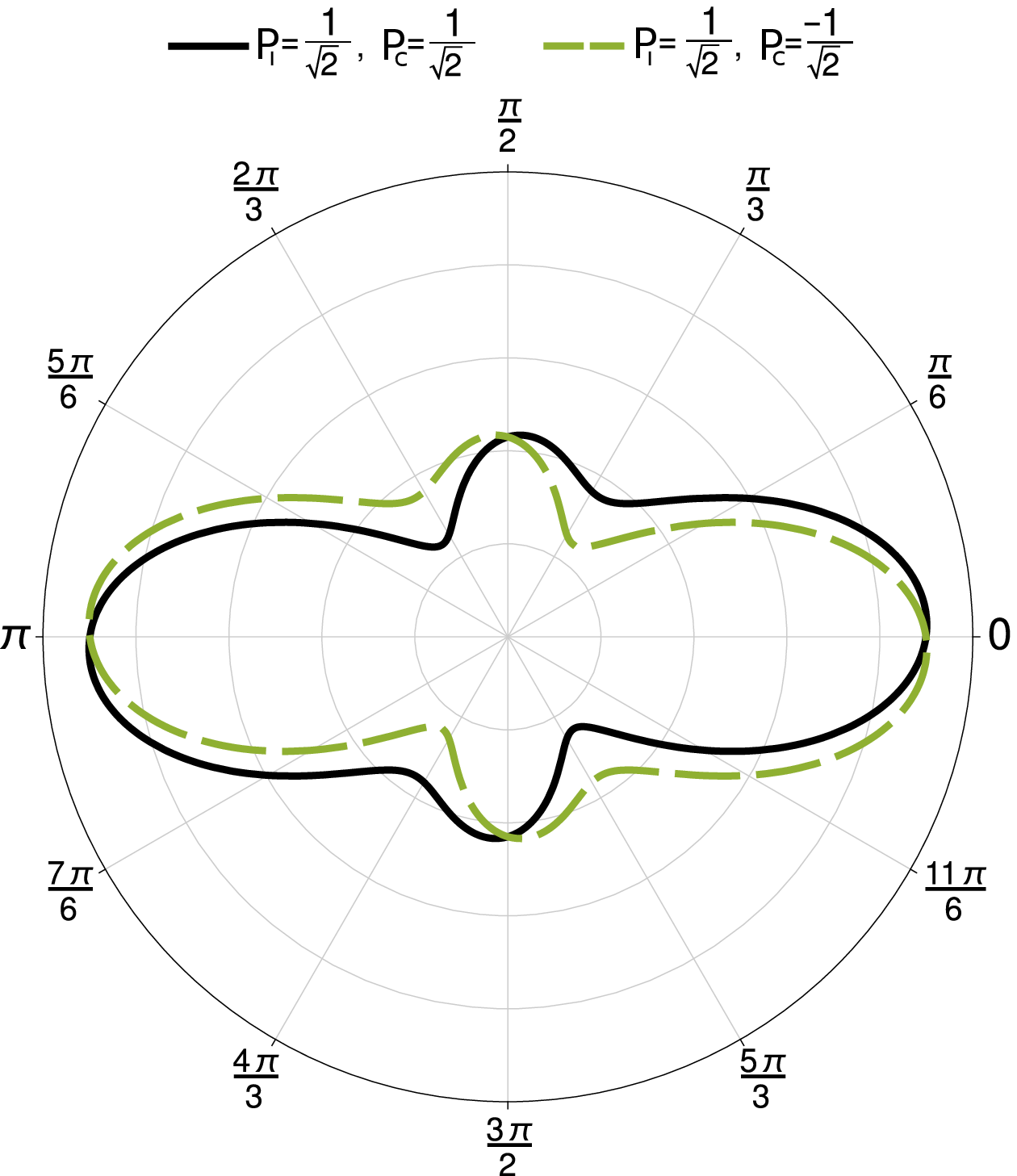}
\endminipage
\caption{Elliptical dichroism in nonresonant two-photon $K$-shell ionization of neutral neon. The angular distributions are presented for the polar angle $\theta=\pi/2$ for two photon excess energies; $\varepsilon=1.05$ (left) and $\varepsilon=1.40$ (right). The sensitivity of the dichroism is clearly visible from the distributions as well as from the relative dichroism parameter $\Delta_{ed}$ (middle).}\label{Fig.Dichorism}
\end{figure*}

From the analysis of relativistic effects, it may seem, that neglecting the screening effects of inactive electrons would not yield a significantly different results, as long as we include higher multipole orders of the photon field. However, from Fig. \ref{Fig.Effects}, it can be seen that this simplification could lead to large errors. This is because the screening effects are significant for nonrelativistic systems, more specifically, for photon energies near the ionization threshold and light elements. The influence of the screening effects was found strongest for the two-photon ionization of neon, where the electron emission along the photon propagation axis exceeds the emission along the linear polarization axis. This result can be best understood in the simplified electric dipole picture, where there are only two ionization channels present; $s$-channel and $d$-channel, where $s$- and $d$- refer to the partial waves of the emitted electron. The contribution of the $s$-channel to the cross section is angle-independent, and is therefore spherically symmetric. On the other hand, the contribution of the $d$-channel is similar to the distribution of two-photon ionization by linearly polarized light with most photoelectrons emitted along the photon polarization direction, see Fig. \ref{Fig.3d}. Generally, the ionization channel with higher angular momentum is dominant, hence, only small contribution to the distribution arises from the $s$-channel. However, in Ref. \cite{Hofbrucker/PRA:2016}, we have shown that due to the screening effects, the dominant $d$-channel drops down for light elements in near-threshold ionization, while the amplitude of the $s$-channel increases. This is exactly what we can see in Fig. \ref{Fig.Effects}. The emission along the polarization axis sharply decreases ($d$-channel contribution) and the emission into all direction slightly increases (the spherical contribution of the $s$-channel). Instead, if we consider the case of ionization by completely circularly polarized light, i.e. $P_c=\pm 1$, only the $d$-channel contributes to the cross section. More specifically, only the $d$-channel with projection of orbital angular momenta $m_e=\pm 2$ contributes. This leads to the typical doughnut shape distribution. Since there is only one active channel, screening effect lead solely to the decrease of the magnitude of the cross section. As none of the described behavior has been observed in two-photon ionization of H-like ions, it must therefore arise from the additional screening potential created by the inactive electrons.

\subsection{Elliptical dichroism}\label{Subsec.Relativistic_effects}

Dichroism of matter represents an asymmetry in the light-matter interaction upon a sign-change of a symmetric property of the light or matter, e.g. handedness of photon polarization or chirality of a molecule. This behavior raises an interest in number of different fields such as material science \cite{Train/NM:2008}, bio-chemistry \cite{Ranjbar/CBDD:2009} and fundamental research \cite{Ilchen/PRL:2017}. Circular dichroism, for example, has been measured in the multiphoton ionization of He ions\cite{Ilchen/PRL:2017} as well as molecular O$_2$ \cite{Hartmann/RSI:2016}. In these experiments, the dichroism arises from the target polarization. We here present the prediction of elliptical dichroism of the photoelectron angular distribution in two-photon ionization process. Unlike the circular dichroism from above references, the elliptical dichroism arises from the interference of ionization paths. In past, elliptical dichroism has been studied theoretically for the multiphoton ionization of outer shell electrons \cite{Manakov/JPB:1999, Wang/PRA:2000}, and was experimentally observed for two-photon ionization of atomic rubidium \cite{Borca/PRL:2001}. The direct analytical origin of the dichroism can be seen in Eq. (\ref{Eq.E1E1}), where only the interference term depends on the handedness of circular polarization. A convenient way to characterise the elliptical dichroism, is by defining the relative dichroism parameter

\begin{eqnarray}
\Delta_{\mathrm{ed}}=\frac{d\sigma_{+}/d\Omega-d\sigma_{-}/d\Omega}{d\sigma_{+}/d\Omega+d\sigma_{-}/d\Omega},
\end{eqnarray}
where the index of $\sigma$ refers to the sign of $P_c$. The dichroism parameter describes the magnitude of the elliptical dichroism and can take maximal values of $\pm 1$. Figure~\ref{Fig.Dichorism} presents the electron distributions for two-photon ionization of Ne at excess energies of $\varepsilon=1.05$ (left) and $\varepsilon=1.40$ (right) as well as the dichroism parameter for both cases (middle). This figure clearly shows the energy dependence of the elliptical dichroism. While for low energies, the difference between $P_c=1/\sqrt{2}$ and $P_c=-1/\sqrt{2}$ is large and the dichroism parameter nearly reaches unity for four given values of $\phi$, the dichroism for the higher energy is much weaker. A similar behavior applies to the dependence on nuclear charge, with the dichroism being strongest for Ne atom. This happens again due to the screening effects. Since the dominant $d-$ ionization channel decreases, and $s-$channel increases, the relative amplitude of the interference term of Eq. (\ref{Eq.E1E1}) increases, and hence, the sensitivity to the handedness of light increases. As shown in Ref. \cite{Hofbrucker/PRA:2016}, the screening effects are strongest for neon, therefore, even the dichroism is largest for ionization of the neon atom. Since the screening effects are fully encapsulated in the transition amplitudes and since the relativistic effects are low for light elements, the elliptical dichroism can be described well with the nonrelativistic expression (\ref{Eq.E1E1}).

Let us briefly discuss the experimental possibility of detecting the elliptical dichroism. With photon energies in the order of keV, today's free electron facilities reach (and exceed) the energy limits to ionize a $K$-shell electron of light atoms. While early free electron lasers were restricted to linear beam polarization, recently, the photon polarization of these beams has been controlled. The polarization control has been achieved, for example, by the Delta undulator at LCLS \cite{Lutman/NP:2016} or Apple II at FERMI \cite{Allaria/NP:2013}. While the Delta undulator operates at 0.5-1.2~keV, the Apple II can produce photon energies up to 120~eV. For detecting the elliptical dichroism, the two-photon energy should be close to the ionization threshold. In practice, therefore, slow photoelectron would be produced from the two-photon $K$-shell ionization and fast photoelectrons from other processes.  The slow electrons could be successively detected by a angle-resolving time-of-flight spectrometer. Similar angle-resolved studies of multiphoton ionization processes at free electron facilities have been carried out for example in Refs. \cite{Hartmann/RSI:2016, Ilchen/PRL:2017}. However, although these experimental conditions seem to be fulfilled, the ionization yields for non-resonant two-photon $K$-shell ionization at photon energies in the orders of few hundred eV are very low, making it challenging to measure the dichroism. However, the dichroism is not a unique feature of $K$-shell two-photon ionization. It will be the subject of further initiative to study two-photon ionization of higher electronic shells, for which the experiment would be more feasible.

%
%
%
%
%
\section{Summary and outlook}
\label{Sec.SummaryAndOutlook}

We have theoretically studied the photoelectron angular distribution of the two-photon ionization of neutral atoms. We presented an exact relativistic expression, as well as nonrelativistic expression in electric dipole approximation for the angle-differential cross section. The latter fully characterizes the photoelectron angular distribution by two parameters; transition amplitude ratio and phase difference of the two dominant channels. Unlike the nonrelativistic cross section of Ref. \cite{Manakov/JETP:2009}, our nonrelativistic cross section is applicable for any degree of photon polarization. We have shown that the nonrelativistic expression is insufficient for describing two-photon ionization by high energetic photons. In such cases, the relativistic effect become important, and photoelectron emission into the forward direction dominates. For photon energies on the other side of the spectra, i.e. near ionization threshold, screening effects become important. For light atoms, the electron screening significantly influences the photoelectron angular distributions, and it leads to a strong elliptical dichroism. As this behavior arises from the transition amplitude ratios and corresponding phases, it can be therefore fully described by the nonrelativistic formula. For an appropriate choice of an atomic system, the elliptical dichroism could be experimentally verified by carrying out a complete experiment.

\begin{acknowledgments}
The authors thank Dr. Markus Ilchen for helpful discussions. This work has been supported by the BMBF (Grant No. 05P15SJFAA).
\end{acknowledgments}

\end{document}